\documentclass[twocolumn,pre,showpacs,preprintnumbers]{revtex4}

\usepackage{graphicx}% Include figure files
\usepackage{bm}% bold math
\newcommand{\bgamma}{\mbox{\boldmath{$\gamma$}}}
\newcommand{\ba}{\mbox{\boldmath{$a$}}}
\newcommand{\bA}{\mbox{\boldmath{$A$}}}

\newcommand{\bomega}{\mbox{\boldmath{$\omega$}}}
\newcommand{\bphi}{\mbox{\boldmath{$\phi$}}}

\newcommand{\bOmega}{\mbox{\boldmath{$\Omega$}}}
\newcommand{\bGamma}{\mbox{\boldmath{$\Gamma$}}}
\newcommand{\bO}{\mbox{\boldmath{$O$}}}
\newcommand{\bW}{\mbox{\boldmath{$W$}}}

\newcommand{\bS}{\mbox{\boldmath{$S$}}}

\newcommand{\bB}{\mbox{\boldmath{$B$}}}

\begin{document}

\title{{\rm Published as: Physica A 348, 611-629 (2005)} \vspace{1cm}\\ A renormalization group theory of cultural evolution}

\author{G\'abor F\'ath$^{1,2}$ and Miklos Sarvary$^2$}

\affiliation{$^1$Research Institute for Solid State Physics and Optics, P.O. Box 49, H-1525 Budapest, Hungary \\
$^2$INSEAD, Boulevard de Constance, 77305, Fontainebleau, France}

%\date{\today}

\begin{abstract}
We present a theory of cultural evolution based upon a renormalization group scheme. We consider rational but cognitively limited agents who optimize their decision making process by iteratively updating and refining the mental representation of their natural and social environment. These representations are built around the most important degrees of freedom of their world. Cultural coherence among agents is defined as the overlap of mental representations and is characterized using an adequate order parameter. As the importance of social interactions increases or agents become more intelligent, we observe and quantify a series of dynamic phase transitions by which cultural coherence advances in the society. A similar phase transition may explain the so-called ``cultural explosion" in human evolution some 50,000 years ago.
\end{abstract}

% PACS, the Physics and Astronomy Classification Scheme.
%\keywords{cultural evolution, bounded rationality, renormalization group}%Use showkeys class option if keyword display desired
\maketitle

\section{Introduction}

Culture is the sum of knowledge, beliefs, values and behavioral patterns built up by a group of human beings and transmitted from one generation to the next. Cultural evolution for \emph{Homo sapiens} started early in the prehistoric age, and as archeological evidence suggests it has gone through a number of short episodes during which its prevalence and overall influence on everyday life increased abruptly. The most important of these events  occurred approximately 50,000 years ago, giving birth to art, music, religion and warfare. This episode is usually termed the ``cultural explosion" \cite{Mithen1996, Klein2002}.

The evolution of human culture is investigated in many disciplines using different paradigms. Economics uses game theory to understand the development of rules, social norms and other cultural institutions assuming rational behavior of the constituting individuals (agents) \cite{Simon1997}. Cognitive and behavioral sciences study how culture is represented in the mind \cite{Romney1996}, and how these ``mental representations" change under social interactions \cite{Aunger2001}. Physics views culture and its evolution as a complex, dynamic system, and aims to identify its basic universal properties using simple minimal models \cite{Deffuant2000,Axelrod1997cikk,Axelrod1997book,Castellano2000,Klemm2003,Weisbuch2004}.

The emergence of culture seems to involve a chicken-and-egg problem: culture becomes what the constituting agents make it to be, and agents, based on their capacity of learning, constantly adapt to the culture they happen to live in. In this sense culture is the dynamic attractor of the complex social dynamics, determined by the agents' actual constraints and social interactions. The aim of this paper is to investigate the properties of such cultural attractors in an adequately formulated social dynamics model, which integrates the relevant concepts from the social sciences with those from the physics perspective.

One of the popular minimal models of cultural evolution is that of Axelrod \cite{Axelrod1997cikk,Axelrod1997book}. Axelrod's model investigates an initially random population of biologically and economically uniform agents. Social interaction is limited to an imitation process in which agents adapt cultural traits stochastically from each other with a bias toward already similar agents. The interesting question is under what circumstances global cultural diversity gets preserved despite the homogenizing interaction \cite{Deffuant2000,Castellano2000,Klemm2003,Weisbuch2004}. However, the question can also be posed from the opposite direction: in a population with non-negligible heterogeneity and under social interactions which are not necessarily imitational but rather reflect selfish individual interests how can a \emph{coherent} culture emerge?

In order to study this question, our model differs from that of Axelrod in two ways. First, we explicitly consider individual decision making mechanisms and identify ``dimension reduction" as a fundamental heuristic for cognitively limited but otherwise rational agents. This is one of the possible implementations of the concept of \emph{bounded rationality} well-known in economics and social sciences \cite{Simon1997,Conlisk1996}. Specifically, agents are constrained to describe and understand their world along a finite number of ``concepts" (Concepts). The totality of these Concepts constitutes the agent's cultural profile. The Concepts establish an interface to the objective world: agents use them to build an adequate mental representation of natural and social reality for evaluating decision alternatives. Obviously, decision making is successful if evaluation is precise, thus rational agents face the problem of optimizing their Concepts given the actual state of the world. Innate or acquired heterogeneity in individual preferences, however, can give rise to substantial deviations in these optimal Concepts across members of the population. People with such differences in cultural profile end up having unequal views of the world, they evaluate alternatives differently, and make different choices accordingly \footnote{The idea that concepts one uses influence her thinking and decisions dates back to E. Sapir and B. L. Whorf (Sapir-Whorf hypothesis) \cite{Carroll1956}. See also Ref. \cite{Ross2004} for a more recent perspective on the issue.}.

Second, instead of a simple imitation process our agents are assumed to interact in a competitive game \cite{Fudenberg1991}. We think of the Concepts as flexible mental constructs which are continuously updated and refined according to a ``best response dynamics" in order to maximize the agents' predictive power in the actual natural and social environment. As such, our model treats the evolution of cultural profiles (the set of Concepts) as the result of rational behavior governed by economic benefit. The Concepts are continuous variables and agents are assumed heterogeneous in their fixed preferences by default.

Conceptually, our theory will be a renormalization group theory as it is based on the following two key elements: (i) it combines the original state variables (attributes of decision alternatives) into important and less important degrees of freedom; agents keep the former (defined to be the Concepts), and discard the latter; (ii) this is done in an iterative way taking into account the other agents' profiles and aiming to identify a fixed point (associated with Culture). The criterion for truncating the degrees of freedom is local: each agent tries to maximize its evaluation accuracy, i.e., to minimize the representation/evaluation error within its cognitive limits.

\section{A cognitive model}

In order to define a simple, tractable model, we will introduce two simplifying assumptions: we will assume that the objective world in which the agents live is \emph{linear}, and that the subjective heuristics that they use to predict their environment is also \emph{linear}, in a sense to be made precise below. Clearly, these are important simplifications as both the real world and the real mental models are likely to exhibit substantial non-linearities. For a first analysis, however, this approximation is sufficient.

We consider $I$ agents, each restricted to use a number $K$ of Concepts only. Agents divide the world into a number $X$ of {\em contexts} in which they evaluate decision alternatives according to their personal preferences. We assume that alternatives are characterized by their objective (physical) attributes $\bm{a}=\{a_1,\ldots,a_D\}$. Agent $i$'s {\em theoretical payoff} from choosing alternative $\bm{a}$ in context $x$ is posited to be a linear function (``linear world assumption")
\begin{equation}
  \pi_i^{(x)}(\ba) = \bomega_i^{(x)}\cdot\ba ,
  \label{objutil}
\end{equation}
where $\bomega_i^{(x)}$ is the agent's {\em preference vector}, characterizing the agent's personal (biological or otherwise acquired) fixed preference in context $x$. For each agent there are $X$ preference vectors each of dimension $D$, which are assumed fixed in the model. %, i.e., altogether $XD$ parameters, all assumed fixed.

An agent does not know his preference vectors explicitly as this would require a detailed understanding of the effect of all attributes on his payoffs. However, by collecting experience on choices he has made previously, he learns to approximate the payoffs using an appropriate \emph{mental representation}. The mental representation is built around the world's $K$ ``most important degrees of freedom", constituting the agent's Concepts. We assume again that the {\em approximate payoff} that the agent ``computes" directly is linear (``linear representation assumption")
\begin{eqnarray}
  \tilde{\pi}_i^{(x)}(\ba) = \tilde{\bomega}_i^{(x)}\cdot\ba
  \label{approxutil}
\end{eqnarray}
with
\begin{eqnarray}
  \tilde{\bomega}_i^{(x)} =
  \sum_{\mu=1}^K v_{i\mu}^{(x)} \bgamma_{i\mu},
  \label{approxutil2}
\end{eqnarray}
the agent's {\em approximate preference vector} (that the agent knows explicitly). $\tilde{\bomega}_i^{(x)}$ is decomposed using \emph{mental weights} $v_{i\mu}^{(x)}$ in a reduced subspace of dimension $K$. The weight $v_{i\mu}^{(x)}$ reflects the significance of Concept $\mu$ in context $x$. Equation (\ref{approxutil}) implies that agent $i$ possesses a number $K$ of concept vectors, $\{\bgamma_{i\mu}\}_{\mu=1}^K$, assumed normalized $|\bgamma_{i\mu}|=1$, which the agent uses to evaluate alternatives. We emphasize that the Concepts are defined in a context-independent way, i.e, they have the same ``meaning" in all occurring contexts.

%%%%%%%%%%%%%%%%%%%
\begin{figure}[tbp]
\centerline{\includegraphics[scale=.55]{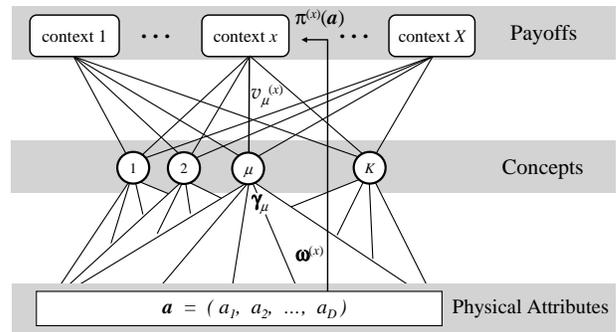}} %preprint:.8, twocolumn:.55
\caption{Decision making heuristic using Concepts as renormalized degrees of freedom.}
\label{fig:3layer}
\end{figure}
%%%%%%%%%%%%%%%%%%%

The structure of our model is depicted in Fig. \ref{fig:3layer}. We have defined three layers: the layer of ``Physical Attributes" of the decision alternatives, the layer of Concepts (defined by concept vectors) which constitute the renormalized degrees of freedom, and the layer of context- and agent-specific ``Payoffs". At the lowest level agents are homogeneous and have identical information about alternatives. In the highest layer agents are heterogeneous and have individual payoff functions based on their individual preferences. The middle layer shows partial coherence, whose measure, as we will see, is determined by the strength of social interactions. This is the layer we intend to monitor across society for observing the emergence and evolution of a common Culture.

An example may be enlightening at this point. Think about the world of chess where decision alternatives are the possible moves a player can choose in a given situation. These moves are characterized by the position of the pieces on the board (physical attributes). Calculating the payoff of a move (the probability of winning the game with that move) is basically a two step procedure for human players and chess programs alike. First the move is evaluated along some general Concepts like
``material advantage", ``positional advantage", ``pinned piece", etc. Indeed, research in psychology has demonstrated that a human grand master possesses tens of thousands of such chess-related concepts \cite{Mero1990,Simon1973a}. Second, these concept scores are weighted (mental weights) in an appropriate mental model. The weighting scheme depends on the player's personal skills/style (agent heterogeneity) and on the actual adversary (context dependence). There is a theoretically best move (preference vector) for that player with that adversary in that situation, but that is not known explicitly to the player. Mastering chess amounts to finding an optimal mental representation, i.e., optimal Concepts and optimal weights (approximate preference vectors) which allow a decision close to the theoretical best. In chess there are different ``schools" (cultures) where the applied Concepts have somewhat different meaning. Of course, decision in chess is very non-linear, and our simplified linear model can only grasp the big picture.

The hierarchical structure depicted in Fig.\ \ref{fig:3layer} formally resembles a linear two-layer (concept vectors, mental weights) neural network. Note, however, that it is not a microscopic neural network model of a cognitive function, but a phenomenological (macroscopic) model of a generic decision making strategy.

Under this cognitive architecture the number of variables defining an agent is $(X+D)K$, much less than the total number of parameters describing the world, $XD$. However, there is a price that should be paid for bounded rationality: due to the reduction of dimensionality, $K<D$, the approximate payoff $\tilde{\pi}_i^{(x)}$ deviates from the theoretical payoff $\pi_i^{(x)}$. The agents' goal is to find the best possible set of Concepts and mental weights which minimize the error of the mental representation under the constraint that only $K$ Concepts can be used. The natural measure of agent $i$'s {\em representation error} is the variance\begin{equation}
    E_i = \sum_{x=1}^X \left\langle (\pi_i^{(x)}-\tilde{\pi}_i^{(x)})^2 \right\rangle_x,
\end{equation}
where $\langle .\rangle_x$ is the average over alternatives in context $x$.

We restrict our attention to the case, where the attributes are delta correlated
\begin{equation}\label{deltacorr}
\langle a_d^{(x)}\rangle_x =0,\quad \langle a_d^{(x)}a_{d'}^{(x)} \rangle_x =\delta_{dd'}
\end{equation}
and nontrivial structure is assumed in the agent-specific preference vectors only. This simplifies the forthcoming analysis without losing essential features. With this proviso, the representation error %, i.e., the negative of the agent' overall utility,
turns out to be
\begin{eqnarray}
   E_i = \sum_{x=1}^X
      \left(\bomega_{i}^{(x)} -\sum_{\nu=1}^K v_{i\nu}^{(x)} \bgamma_{i\nu}\right)^2.
   \label{eq:E}
\end{eqnarray}
The agents' goal is to minimize the error $E_i$ by optimally choosing their concept vectors $\bgamma_{i\nu}$ and mental weights $v_{i\nu}^{(x)}$. Recall that $\bomega_{i}^{(x)}$ is assumed fixed in the model.

%%%%%%%%%%%%%%%%%%%%%%%%%%%%%%%%%%%%%%%%%%
\section{Optimal concepts as a PCA problem}

Theoretically the minimizing parameters can be easily determined. Differentiating w.r.t.\ $v_{i\mu}^{(x)}$ and assuming that the $K\times K$ metric tensor $\Gamma_{i\nu\mu}=\bgamma_{i\nu}\cdot \bgamma_{i\mu}$ is invertible (note that the Concepts are not necessarily orthogonal), the optimal mental weights read
\begin{equation}
   v_{i\mu}^{(x)} = \bomega_i^{(x)}\cdot\bgamma_i^\mu,
   \label{mentalweights}
\end{equation}
where we have introduced the ``dual" concept vectors $\bgamma_i^\mu = \sum_{\nu=1}^K [\bGamma_i^{-1}]_{\mu\nu} \bgamma_{i\nu}$. Writing this back to Eq.\ (\ref{eq:E}) we can write the error as
\begin{eqnarray}
   E_i = \sum_{x=1}^X |\bomega_i^{(x)}|^2
    -\sum_{x=1}^X \sum_{\mu=1}^K (\bomega_i^{(x)}\!\cdot\!\bgamma_i^\mu)
    (\bomega_i^{(x)}\!\cdot\!\bgamma_{i\mu}).
   \label{EiL}
\end{eqnarray}
In this way the representation error is written as a function of the $\bgamma$ concept vectors only. Conceptually this is equivalent to the reasonable assumption that the mental weights accommodate much faster than the Concepts themselves.

Neglecting the first term, which is an unimportant constant, we define the agent's
utility as the negative of his overall representation error
\begin{eqnarray}
   U_i &=& \sum_{x=1}^X \sum_{\mu=1}^K (\bomega_i^{(x)}\!\cdot\!\bgamma_i^\mu)
                                       (\bomega_i^{(x)}\!\cdot\!\bgamma_{i\mu}),
   \label{Utility1}
\end{eqnarray}
which should now be maximized for $\{\bgamma_{i\nu}\}_{\nu=1}^K$.
The utility can be written in a more compact form
\begin{eqnarray}
   U_i  &=&
        \sum_{\mu=1}^K \bgamma_{i}^\mu\cdot \bW_{i}\,\bgamma_{i\mu}.
   \label{Utility}
\end{eqnarray}
by introducing a positive semi-definite, $D\times D$ dimensional matrix, to be called the {\em world matrix},
\begin{equation}\label{worldmatrix}
\bW_i= \sum_{x=1}^X \bomega_i^{(x)}\circ\bomega_i^{(x)},
\end{equation}
which encompasses all information about agent $i$'s (objective) personal relationship to the world.

The maximization problem in Eq.\ (\ref{Utility}) is the well-known Principal Component Analysis (PCA) problem. According to this, a particular solution for the optimal Concept vectors is provided by the $K$ most significant (largest eigenvalue) eigenvectors of $\bW_i$. Thus to achieve the best possible mental representation the agent should choose his concept vectors according to the eigenvectors of his world matrix in the order of their significance, and then adapt his mental weights according to Eq.\ (\ref{mentalweights}). Notice, however, that this solution is not unique: any non-singular basis transformation in the subspace of the $K$ most significant eigenvectors yields a different solution with identical utility. We can freely define linear transformations in the subspace of Concepts; the mental weights adapt, and the overall representation error remains unchanged. It is obvious that if $\bW_i$ is a non-singular, rank-$D$ matrix, then for $K<D$ the error is strictly positive even with an optimally chosen set of Concepts.

\subsection*{Relationship with the DMRG}

The identification, sorting and truncation of the degrees of freedom in our model is closely analogous to what occurs in White's Density Matrix Renormalization Group method (DMRG) \cite{White1992} -- a numerical technique widely used in the simulation of strongly correlated quantum systems. In the DMRG the optimally renormalized degrees of freedom turn out to be the $K$ most significant eigenvectors of the reduced density matrix of the quantum subsystem embedded in the environment with which it interacts.

A formal mapping between the DMRG and our method can be established by identifying the DMRG's density matrix for block $i$ with the world matrix $\bW_i$ for agent $i$ (both measuring the subsystem's relationship to the environment), and the renormalized (kept) quantum mechanical degrees of freedom with the Concepts. The DMRG's ``truncation error" is analogous to our representation error. There, the error measures how precisely the entangled quantum mechanical wave function (written in a matrix form) is reconstructed after the truncation of the degrees of freedom; here, the role of the wave function is played by the $D\times X$ dimensional matrix formed by the preference vectors $\bomega_i^{(x)}$. The two problems are conceptually similar in that both seek an optimal linear reduction of dimensionality for the subsystem, which is mathematically equivalent to the PCA problem.

%%%%%%%%%%%%%%%%%%%%%%%%%%%%%%%%%%%%%%%%%%
\section{Social interactions}

As shown above, the representation error is minimal if the agent learns to approximate his world matrix in the $K$ dimensional subspace spanned by the most significant eigenvectors of his world matrix. If agents $i$ and $j$ have different preferences, and thus different world matrices, they would necessarily end up with different optimal Concepts. Cultural evolution boils down to the interaction of these different Concepts in a social network. In order to introduce social interactions we cast contexts into two basic categories: those where preference vectors only depend on a single agent (``individual contexts"), and those where the preference vectors for agent $i$ depend on the (simultaneous) decision/preference of at least another agent $j$ (``social contexts"). For simplicity only pair interactions will be considered. In social contexts the understanding of another agent's valuation and decision making mechanism is an asset. Agent $i$ has strategic advantage from being able to estimate $j$'s payoff for her alternatives. Again, chess is an illustrative example. Always playing against the same unchanging computer software is an individual context, whereas playing against a human adversary capable of learning from her errors is a social context.

In a given context, agent $j$ bases her evaluations on her mental representation which is characterized by her approximate preference vector $\tilde{\bomega}_j^{(x)}$. This is necessarily within her actual Concept subspace. Agent $i$ tries to approximate this within his Concept subspace. The accuracy of this approximation depends on the overlap of the two subspaces. Thus social contexts introduce a ``force" which tries to deform $i$'s Concept subspace towards that of $j$ in order to improve $i$'s predictive power in $j$-related social contexts. The interplay between the two competing goals -- optimizing the mental representation for social contexts vs.\ individual ones -- is the ultimate factor which determines the society's cultural coherence.

In terms of the above classification, agent $i$'s overall world matrix is a sum of individual and social contributions. As seen above, individual contexts contribute
\begin{equation}
    \bW^0_i=\sum_{x=1}^X \bomega_{i}^{(x)}\circ\bomega_{i}^{(x)};\quad x=\mbox{individual},
\end{equation}
and the contribution of all $j$-related social contexts is now
\begin{equation}
    \bW^j_i=\sum_{x}\tilde{\bomega}_j^{(x)}\circ \tilde{\bomega}_j^{(x)}
    \quad x=\mbox{$j$-related social}.
\end{equation}
The interpretation of $\bW^j_i$ is that agent $j$'s approximate preference vectors (which determine her decisions) are fully presented to agent $i$ (they build into his world matrix), but -- in accordance with the central thesis of our model -- agent $i$ can only pick up from this what his internal Concepts span.

We realize that $\bW^j_i$ is a rank-$K$ operator, which can be expressed explicitly in terms of $j$'s concept vectors $\{\bgamma_{j\mu}\}_{\mu=1}^K$ and mental weights $v_{j\mu}^{(x)}$ using Eq.\ (\ref{approxutil2}). This explicit representation has the structure
\begin{equation}
    \bW^j_i=\bB\, \bS_j,
\end{equation}
i.e., the product of
\begin{equation}
    \bS_j=\sum_{\mu=1}^K \bgamma_{j\mu}\circ\bgamma_{j}^\mu\;,
\end{equation}
projecting onto agent $j$'s Concept subspace, and another operator $\bB$ acting within this subspace. This latter depends on the actual $v_{j\mu}^{(x)}$ weights, but for simplicity in what follows we assume that it is proportional to the identity operator
\begin{equation}
    \bB = h_{ij}\, \mathbf{1}.
\end{equation}
This assumption simply states that all directions in $j$'s Concept subspace have \emph{equal importance} for $i$, and is reasonable if $i$ has to predict $j$'s behavior in many different $j$-related contexts which average out the weight dependence.

Eventually, agent $i$'s overall world matrix to be used in the utility Eq.\ (\ref{Utility}) takes the form
\begin{equation}
    \bW_i = \bW^0_i + \sum_{j\in {\cal N}_i} h_{ij}\, \bS_j,
    \label{culture1}
\end{equation}
the sum being over the agent's social network ${\cal N}_i$. The parameters $h_{ij}$ measure the relative strength (importance) of social interactions with agent $j$. The equal importance assumption assures that our $\bW_i$ remains invariant w.r.t.\ nonsingular local linear transformations of the individual concept vectors. This means that in our model the actual choice of Concepts do not count, only the subspace they span do.

Using Eq.\ (\ref{culture1}) in Eq.\ (\ref{Utility}) allows us to write the agent's utility function in an explicit form, which only depends on the concept vectors,
\begin{equation}\label{explicitutility}
    U_i = \sum_{\mu=1}^K \bgamma_{i}^\mu\cdot \bW^0_{i}\,\bgamma_{i\mu}
          + \sum_{j\in {\cal N}_i} h_{ij}\, \sum_{\mu,\nu=1}^K
          (\bgamma_{i}^\nu\cdot\bgamma_{j\mu})(\bgamma_{i\nu}\cdot\bgamma_j^\mu).
\end{equation}

\subsection*{Adjustment dynamics}

The introduction of social interactions in our model of mental representations results in an economic game where the agents' optimization problems are potentially in conflict.
We can view the society as a dynamical system in which agents continuously act as  best-response optimizers \cite{Hofbauer1998}, in each step maximizing their utility given the actual state of nature and other agents (\emph{best-response dynamics}). Alternatively, and this is more consistent with bounded rationality, we can postulate a slow, continuous adjustment dynamics which drives agents towards ever better responses. In this spirit the time evolution of the concept vectors is defined to be
\begin{eqnarray}\label{adjust}
   \frac{\delta \bgamma_{i\mu}}{\delta t}=\mbox{const}\,
   \frac{\partial U_i}{\partial \bgamma_{i\mu}},
\end{eqnarray}
where $U_i$ is given by Eq.\ (\ref{explicitutility}) (\emph{gradient adjustment dynamics}). Since in the general case $U_i$ depends on all agents' all Concepts, this in fact amounts to $I\times K$ coupled equations. As we will see below the two dynamics lead to the same result.

\subsection*{The cultural coherence order parameter}

In order to monitor cultural coherence we introduce a \emph{coherence order parameter} (COP) as the population average of Concept subspace projector operators
\begin{eqnarray}
   \bO = \left\langle \bS_j \right\rangle_I=
         \left\langle \sum_{\mu=1}^K \bgamma_{j\mu} \circ \bgamma_j^\mu \right\rangle_I.
       \label{OP}
\end{eqnarray}
Our COP is a tensor order parameter which measures the average overlap of the individual Concept subspaces. In fact it is a high-dimensional generalization of de Gennes's tensor order parameter introduced in the theory of liquid crystals \cite{deGennes1974}. The eigenvalue structure of $\bO$ is useful to characterize the level of coherence in the society and to distinguish different phases. Obviously, if there
is \emph{perfect order} and the individual Concept subspaces are all lined up, the COP has eigenvalues $o_1=o_2=\dots=o_K=1$, and $o_{K+1}=o_{K+2}=\dots=o_D=0$. In the opposite extreme of \emph{complete disorder} all directions are equivalent and the eigenvalues are $o_1=o_2=\dots=o_D=K/D$. The trace of the COP is always $K$.

%%%%%%%%%%%%%%%%%%%%%%%%%%%%%%%%%%%%%%%%%%
\section{Cultural ordering as a phase transition}

The equilibrium and dynamic properties of the social system depend crucially on the parameters $D,X,K$, the statistical properties of $\bW^0_i$, and the social interaction matrix $h_{ij}$ (connectivity structure). In the following we analyze the case when the agents' individual preference vectors $\bomega_i^{(x)}$ are \emph{Gaussian random vectors} with zero mean and unit variance, i.e., without interaction agents choose random Concept subspaces (complete disorder). This is a benchmark case which can demonstrate in its purest form how culture can emerge spontaneously in an interacting social system. With this assumption $\bW^0_i$, which is quadratic in the preference vectors, has Wishart distribution \cite{Wishart1928,Edelman1989}. We restrict our attention to a \emph{mean-field} network (all agents are connected) and set $h_{ij}=h/I$. With this the world matrix simplifies to
\begin{equation}\label{MFWi}
    \bW_i = \bW^0_i + h\, \bO.
\end{equation}

As discussed above there are various possibilities for reasonable evolutionary dynamics. We have analyzed the social behavior under two: the best response dynamics and the continuous gradient adjustment dynamics defined in Eq.\ (\ref{adjust}). In the best response dynamics we started from a certain initial condition of individual Concepts. In each step, we first calculated the actual COP, then updated (with a certain probability to make the update asynchronous) each agent's Concepts according to the $K$ largest principal components of $\bW_i$. This gave agent $i$'s best response for the actual environment (individual plus social). In the gradient adjustment method we discretized the differential equation in Eq. (\ref{adjust}).

Fixed points of the best response dynamics are Nash equilibria, where agents have no incentive to unilaterally deform their mental representations, because these are already mutually optimal. Fixed points attained dynamically by gradient adjustment, in turn, are necessarily stable against small collective perturbations in the concept vectors, i.e., when several agents consider slight deviations simultaneously, but not necessarily optimal on a global scale. It is a special feature of our model (which is usually not the case in other models) that these two dynamics have identical fixed points, i.e., the attained fixed point of the gradient dynamics is also a Nash equilibrium, and vice versa. This feature stems from the fact that the landscape of the PCA's quadratic error function is known to be smooth with a unique minimizing subspace \cite{Baldi1995}. This is enough to assure the equivalence of fixed points. However, in the case of multiple agents (coupled PCA problems) the number of fixed points may not be unique. In fact our results demonstrate that it is typical to have many equilibria, the structure of which depends strongly on $h$.

In case of multiple equilibria, equilibrium selection becomes important. In the following we analyze a situation when the strength of social interactions $h$ is varied slowly. An adiabatic change in the exogenous parameters assures that the fixed point the system has reached is tracked analytically, except for possible bifurcations. We start with zero coupling, $h=0$, where the (unique) fixed point arises as the collection of individual PCA solutions.

%%%%%%%%%%%%%%%%%%%
\begin{figure}[tbp]
\centerline{\includegraphics[scale=.47]{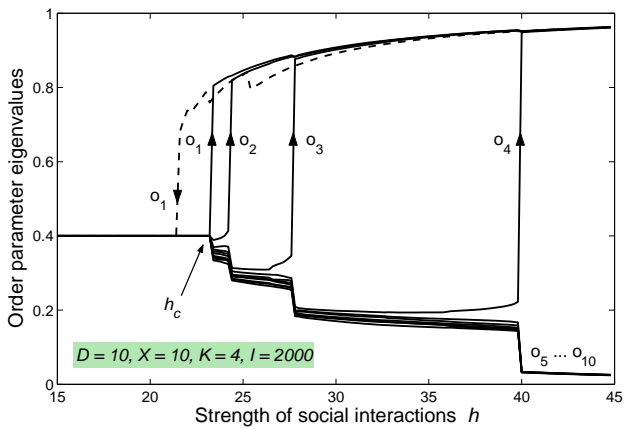}} %preprint:.65, twocolumn:.47
\caption{Fixed point spectrum of the order parameter matrix $\bO$ for $D=X=10$, $K=4$, $I=2000$. Solid lines apply for increasing $h$, dashed line (only shown for $o_1$) for decreasing $h$. The phase transitions occur as subcritical bifurcations.}
\label{fig:K4}
\end{figure}
%%%%%%%%%%%%%%%%%%%

The COP spectrum in the fixed point, determined by a numerical simulation with adiabatic increase (and decrease) of $h$ for $D=10$, $X=10$, $K=4$ and $I=2000$ is depicted in Fig.\ \ref{fig:K4}. The finite population of agents with random world matrices were artificially ``symmetrized" to assure $\langle\bW^0_i\rangle=K/D\, \mathbf{1}$ exactly even in a finite sample at $h=0$. This symmetrization procedure, however, is irrelevant if the sample is large enough.

According to the figure, for small $h$ there is complete disorder, and all eigenvalues are equal, $o_k=K/D$, $\forall k$. In other words there is no systematic overlap between the agents' Concepts. For $h\to\infty$ the fixed point is completely ordered: $o_k=1$ if $k\le K$, zero otherwise, meaning that agents share all $K$ Concepts, i.e., we can talk about a coherent culture. In between we observe a series of dynamic phase transitions, each associated with the sudden emergence of an additional shared Concept. Each transition can be associated with a first order jump in the eigenvalues through a subcritical bifurcation. In accordance with the subcritical nature of the bifurcation scheme hysteresis is observed. Figure \ref{fig:K4} only shows how $o_1$ behaves as $h$ diminishes from its maximum value, but the behavior is similar for the other eigenvalues too. A qualitatively identical picture, with a series of first order transitions and hysteresis, was found numerically for other parameter values, as well.

Our results suggest that the social dynamics has a unique equilibrium only for small couplings. Above the critical value $h_c$ spontaneous ordering occurs in the system, and the number of equilibria increases. On the one hand, there is a discrete number of possible fixed point ``families" as represented by the eigenvalue structure of the COP: an adiabatically slow change in $h$ leads to bifurcations at some critical values, where the system jumps from one equilibrium to another. These transitions are irreversible and give rise to hysteresis, which assures the coexistence of at least two different equilibrium families within the hysteresis loop. On the other hand, each family of equilibria is infinitely degenerate in itself, since the associated \emph{eigenvectors} of the COP can point in essentially any direction (respecting orthogonality). Since we assumed that preferences are random, there are no preferred directions on the social level by default. This symmetry of the $D$-space is broken spontaneously above $h_c$. As $h\to\infty$ there is only one possible structure for the eigenvalues of the COP (there is only one family), meaning that the society is fully coherent. However, this coherent $K$-dimensional subspace, representing the most important dimensions of the society is not fixed a priori. The ultimate state of culture gets selected as a result of the idiosyncratic fluctuations present at the time of the symmetry breaking bifurcations.

%%%%%%%%%%%%%%%%%%%%%%%%%%%%%%%%%%%%%%%%%%
\subsection*{Calculating the critical point}

The critical interaction strength $h_c$, where the completely disordered phase loses stability for increasing $h$ can be calculated analytically for $I\to\infty$. Starting from the disordered phase we can write the COP in step $l$ as
\begin{equation}\label{Oiter}
\bO=(K/D)\mathbf{1} + \varepsilon\, \delta\bO_l,
\end{equation}
where $\delta\bO_l$ is an arbitrary perturbation. The dynamics defines a mapping
\begin{equation}\label{dOmapping}
\delta\bO_l\to\delta\bO_{l+1}.
\end{equation}
If the disordered state is stable we have $||\delta\bO_l||\to 0$ as $l\to\infty$, otherwise $||\delta\bO_l||$ diverges. We can identify $h_c$ as a fixed point of this mapping
\begin{equation}\label{dOfixedpoint}
\delta\bO_{l+1}=\delta\bO_l,
\end{equation}
and determine $\delta\bO_{l+1}$ as a function of $\delta\bO_{l}$ and $h$ using first order perturbation theory in $\varepsilon$. The calculation necessarily involves the spectral properties of the random $\bW^0_i$ ensemble, and leads to (see the Appendix for details)
\begin{eqnarray}
   h_c = \frac{D^2}{2\sum_{\nu=1}^K \xi_\nu},
   \label{hc1a}
\end{eqnarray}
with
\begin{eqnarray}
   \xi_\nu = \left\langle \sum_{m\ne\nu}^D \frac{1}{\lambda_{i\nu}-\lambda_{im}}  \right\rangle_I ,
   \label{hc1b}
\end{eqnarray}
where $\lambda_{im}$ are eigenvalues (in descending order) of the Wishart matrix $\bW^0_i$, and $\nu$ is the index for Concepts.

When $\nu<<D,X$ we find that $\xi_\nu$ is $\nu$-independent, and takes the form (see the Appendix)
\begin{equation}\label{xi_nu}
    \xi_\nu = \frac{1}{2}\left( 1-\frac{(X-D-1)}{D+X+2\sqrt{DX}} \right).
\end{equation}
In the limit $D,X\to\infty$ with $X/D=r>0$ and $K<<D$ this leads to the critical coupling
\begin{eqnarray}
   h_c \approx \frac{D^2}{K}\;\frac{1+\sqrt{r}}{2}.
   \label{hc2text}
\end{eqnarray}

\begin{figure}[tbp]
\centerline{\includegraphics[scale=.5]{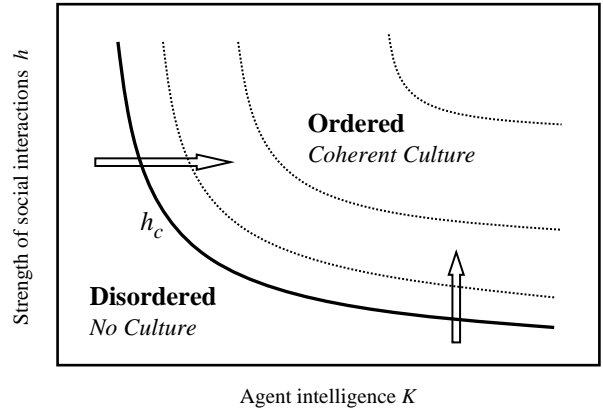}} %preprint:.7, twocolumn:.5
\caption{Schematic phase diagram on the $h$ vs.\ $K$ plain ($D,X=\mbox{fixed}$) for the mean-field model with random preferences. Solid line indicates the principal transition calculated in the text, dashed lines the additional transitions.}
\label{fig:phdiag}
\end{figure}

Keeping $r$ fixed, $h_c$ increases in $D$ (quadratically) and decreases in $K$. When the World is complex (large $D$ and $X$) and the agents are primitive (small $K$) the critical coupling is high and the society is likely to stay culturally disordered. A schematic phase diagram on the $h$ vs $K$ plain is presented in Fig.\ \ref{fig:phdiag}. The $h_c$ curve cuts the plain into two basic domains: one which is disordered and one which is spontaneously ordered. This latter is divided further into sub-domains with different levels of order, separated by first order transition lines.

%%%%%%%%%%%%%%%%%%%
%%%%%%%%%%%%%%%%%%%

%%%%%%%%%%%%%%%%%%%%%%%%%%%%%%%%%%%%%%%%%%
\section{Conclusions}

In summary, in this paper we have proposed a theory of cultural evolution built upon a renormalization group scheme. We assumed that bounded rationality forces agents to choose a limited set of concepts in describing their world. These concepts and the way they are used are continuously updated and refined in order to find an optimal representation. Individual interest in making the best possible decisions in natural (individual) and social (collective) contexts act as a driving force for cultural evolution and, at least in the particular cases studied, drive the system towards a Nash equilibrium fixed point with measurable level of social (cultural) coherence. In our model with random preferences this coherence manifested itself as a spontaneous ordering of the individual concept (sub)spaces, and was quantified by an adequate order parameter. We demonstrated that as the fundamental cognitive and social parameters change, cultural  evolution goes through a number of phase transitions, which we found first order in the mean-field model. The critical value of the most relevant transition was calculated analytically using linear stability arguments and random matrix theory.

Transition to a phase with substantially higher cultural coherence can be triggered in two different ways (see the arrows in Fig.\ \ref{fig:phdiag}): social contexts can increase in significance, or alternatively, agents may develop better cognitive skills. In any case when one of the transition lines is crossed, the subspaces spanned by the agents' concepts rearrange into a state with higher average overlap, and we can speak about a boost in cultural coherence. A similar transition may explain the sudden cultural explosions for humanity approximately 50,000 years ago, and a further jump about 10,000 years ago in connection with the emergence of agriculture. Since in our model the ordering involves a spontaneous symmetry breaking, the emerging fixed point strongly depends on the initial perturbation (a phenomenon usually termed ``path dependence" in the social sciences), implying that cultural evolution is not completely deterministic but depends on a number of idiosyncratic factors.

Our theory differs conceptually from some traditional thinking on the possible causes underlying human cultural explosion. Most of these traditional explanations presume a sudden and substantial \emph{biological} change (genetic mutations \cite{Ding2002,Klein2002}, the integration of cognitive modules in the brain \cite{Mithen1996}, etc.) as the initial trigger. In contrast with these, the mechanism we have proposed in this paper suggests a \emph{spontaneous ordering} of the mental representations, which emerges when \emph{slow changes} in some relevant variables (cognitive abilities, social interactions) reach a critical level. In this respect the ``cultureless state" of the society, originally rooted in the heterogeneity of its members in individual preferences, needs, experiences and the extreme simplicity of the mental representations they possess, becomes unstable at a bifurcation point, beyond which the collective social dynamics drives the system into a drastically different, culturally ordered fixed point.

It would be interesting to go beyond mean field, and investigate the model's dynamics in more realistic social networks, where the fixed point may possess a more complex internal structure. A non-mean-field-like social structure is expected to give rise to subcultures or cultural domains, where individuals are highly coherent within a domain, but the average cultural subspace the domain represents may be rather different from that of another domain, as occurs usually in real societies. The type of the occurring dynamic transitions may also change character. An evaluation of how a potential structure in the attribute space or non-linearity in the world/mental representations influence the results could also be interesting. Other intriguing problems involve the social dynamics itself, namely how the model should be extended to cope with intrinsically dynamical aspects of culture such as fads and fashions, where the social attractor is much rather a limit cycle or a chaotic attractor.

\begin{acknowledgments}
The authors acknowledge support from INSEAD Foundation and the Hungarian Scientific Research Found (OTKA) under grant Nos. F31949, T43330 and T47003.
\end{acknowledgments}

\appendix*
%%%%%%%%%%%%%%%%%%%%%%%%%%%%%%%%%%%%%%%%%%
\section{}

In the disordered phase the order parameter is diagonal
\begin{equation}
   \bO={K\over D}{\bf 1}.
\end{equation}
In order to assess the stability of this phase assume that in step $l$ of the
iterated dynamics the actual order parameter differs from this by a small perturbation
\begin{equation}
   \bO^{\rm ini}={K\over D}{\bf 1} + \varepsilon O_l,
\end{equation}
where $\varepsilon$ is infinitesimal.
In the next step, using $\bO^{\rm ini}$, we should first create the agent specific
context matrices
\begin{equation}
   \bW_i =\bW_i^0 + h \bO^{\rm ini},
\end{equation}
and diagonalize these to obtain the actual (orthogonal) concept vectors
\begin{equation}\label{Wieigen}
   \bW_i \bgamma_{in} = \eta_{in} \bgamma_{in}, \quad n=1,\dots,D.
\end{equation}
When the eigenvalues $\eta_{in}$ are sorted in descending order, the agent keeps Concepts $n=1,\dots,K$.
To finish this iteration step we calculate the updated (final) order parameter as
\begin{eqnarray}
   \bO^{\rm fin} = \left\langle \sum_{\nu=1}^K
                   \bgamma_{i\nu}\circ \bgamma_{i\nu} \right\rangle
               = {K\over D}{\bf 1} + \varepsilon \bO_{l+1}.
               \label{Ofin}
\end{eqnarray}
which defines $\bO_{l+1}$.
The disordered state is stable and the perturbation dies away if $||\bO_l||\to 0$ as $l\to\infty$; otherwise it is unstable.

In fact each step of the dynamics defines a mapping from $\bO_l$ to $\bO_{l+1}$.
This mapping can be approximated as linear if $\varepsilon$ is small. $\bO_l$ can be decomposed into $D\times D$ ``eigenmatrices" each of which transforms into a constant (the eigenvalue) times itself under the mapping. The disordered state loses stability when the maximal eigenvalue becomes greater than one. Thus $h_c$ can be found by seeking the solution of the equation
\begin{equation}
   \bO_{l+1}=\bO_l.
   \label{cond-llp}
\end{equation}

Let us introduce the orthonormal eigenvectors of the Wishart matrix $\bW_i^0$
\begin{equation}
   \bW_i^0 \bphi_{in} = \lambda_{in}\, \bphi_{in}, \quad n=1,\dots,D,
\end{equation}
with $\lambda_{i1}\ge\dots\ge \lambda_{iD}$. When $\varepsilon$ is small we can use {\em first order perturbation theory} to obtain the concept vectors (eigenvectors) in Eq.\ (\ref{Wieigen}). This yields
\begin{eqnarray}
   \gamma_{i\nu} = \bphi_{i\nu} + \varepsilon h\sum_{m\ne \nu}^D
                   \frac{[\bO_l]_{\nu m}}{\lambda_{i\nu}-\lambda_{im}} \bphi_{im}
                   + {\cal O}(\varepsilon^2)
\end{eqnarray}
Substituting this into Eq.\ (\ref{Ofin}) we obtain
\begin{eqnarray}
   \bO^{\rm fin} &=& \left\langle \sum_{\nu=1}^K
                \bphi_{i\nu}\circ \bphi_{i\nu}\right\rangle_I +
                \varepsilon\, h \left\langle
                \sum_{\nu=1}^K\sum_{m=1}^D  \right. \nonumber\\ &&
                \left.\frac{[\bO_l]_{\nu m}}{\lambda_{i\nu} -\lambda_{im}}
                (\bphi_{i\nu}\circ \bphi_{im}+\bphi_{im}\circ \bphi_{i\nu})\right\rangle_I
              + {\cal O}(\varepsilon^2). \nonumber\\
\end{eqnarray}
To evaluate the terms we can use the fact that the eigenvalues and eigenvectors
of the random matrix ensemble $\bW_i^0$ can be considered independent, and
thus the average of a product containing both decouples. Remarking that
\begin{eqnarray}
   \langle \phi_{i\nu d}\phi_{i'\nu' d'} \rangle_I =
   {1\over D} \delta_{ii'}\delta_{\nu\nu'}\delta_{dd'},
\end{eqnarray}
we obtain that the first term is simply $(K/D){\bf 1}$, and the second term defines
$\bO_{l+1}$ as
\begin{eqnarray}
   [O_{l+1}]_{dd'} = h \sum_{m,m'=1}^D A_{dd';mm'}\, [O_{l}]_{mm'}
\end{eqnarray}
where the mapping operator $\bA$ is
\begin{eqnarray}
   A_{dd';mm'} = {1\over D^2} \sum_{\nu=1}^K \xi_\nu \;
                 (\delta_{dm}\delta_{d'm'}+\delta_{dm'}\delta_{d'm})
\end{eqnarray}
with
\begin{eqnarray}\label{xinudef}
   \xi_\nu = \left\langle \sum_{m\ne\nu}^D \frac{1}{\lambda_{i\nu}-\lambda_{im}}  \right\rangle_I
\end{eqnarray}
Using the fact that both $\bO_l$ and $\bO_{l+1}$ is symmetric this reduces to
\begin{eqnarray}
   [O_{l+1}]_{dd'} = \frac{2h}{D^2}\sum_{\nu=1}^K \xi_\nu \, [O_{l}]_{dd'},
\end{eqnarray}
showing that the (normalized) eigenmatrices of the mapping are the $D(D+1)/2$ independent
(unit) matrix elements, and the common eigenvalue is $D(D+1)/2$ times degenerate. In
virtue of Eq.\ (\ref{cond-llp}) this eigenvalue defines the critical social coupling
\begin{eqnarray}
   h_c = \frac{D^2}{2 \sum_{\nu=1}^K \xi_\nu}.
   \label{hc1}
\end{eqnarray}

The only remaining question is how to evaluate the denominator. The theory of random
matrices\cite{Mehta1991} gives a solution. Recall that $\bW\equiv\bW_i^0$ belongs to the {\em real Wishart ensemble} as it is the outer product of a $D\times X$ real
Gaussian matrix with its transpose
\begin{eqnarray}
   \bW = \bOmega\, \bOmega^{\rm T},
\end{eqnarray}
where $\bOmega$ is the $D\times X$ matrix constructed from the
preference vectors $\bomega^{(x)}$ as columns. The joint probability density of eigenvalues for real Wishart matrices
is \cite{Edelman1989,Shen2001}
\begin{eqnarray}
   P(\{\lambda\})\!\!\!\! && d\{\lambda\} = {1\over {\cal Z}}
           \exp\left(-{1\over 2}\sum_{d=1}^{\min(D,X)} \lambda_d \right)\cdot \;
            \nonumber\\ && \prod_{d=1}^{\min(D,X)}\!
            \lambda_d^{(|X-D|-1)/2}
           \!\prod_{n>m}^{\min(D,X)}\! (\lambda_n-\lambda_{m})\,
            d\{\lambda\}\nonumber\\
   \label{jointdens}
\end{eqnarray}
for the ${\min(D,X)}$ nonzero eigenvalues, and there are $D-X$ zero eigenvalues if $D>X$. The partition function ${\cal Z}$ is a normalization constant. It is illuminating to write this in the Coulomb gas representation \cite{Shen2001}
\begin{eqnarray}
   P(\{\lambda\}) &=& {1\over {\cal Z}}
           \exp\left(-\!\!\!\!\sum_{d=1}^{\min(D,X)}\!\!\!\! V(\lambda_d) +
           \!\!\!\!\prod_{n>m}^{\min(D,X)}\!\!\!\! \ln |\lambda_n-\lambda_{m}| \right) \nonumber\\
           \label{Coulombgas}
\end{eqnarray}
with
\begin{eqnarray}
   V(\lambda) = {1\over 2}\lambda -\frac{|X-D|-1}{2}\ln \lambda.
   \label{Vpot}
\end{eqnarray}
In this expression the eigenvalues can be interpreted as coordinates of particles
put in an external potential $V$ and interacting through a 2D Coulomb interaction.
For large $D$ and $X$, the average position of particles (eigenvalues) follows from a {\em saddle-point approximation} leading to the equation
\begin{eqnarray}
   \partial_{\lambda_n} V = \!\!\sum_{m=1}^{\min(D,X)}\!\! {}^\prime \frac{1}{\lambda_n-\lambda_m}.
\end{eqnarray}
Using the explicit form of $V$ in Eq.\ (\ref{Vpot}) this reads
\begin{eqnarray}
   1-\frac{|X-D|-1}{\lambda_n} = 2\!\!\sum_{m=1}^{\min(D,X)}\!\! {}^\prime \frac{1}{\lambda_n-\lambda_m}.
\end{eqnarray}
When the number of zero eigenvalues is taken into account correctly, we obtain for $\xi_\nu$ defined in Eq.\ (\ref{xinudef})
\begin{eqnarray}
   2\xi_\nu = 1-(X-D-1)\left\langle\frac{1}{\lambda_\nu}\right\rangle,
\end{eqnarray}
and thus using Eq.\ (\ref{hc1})
\begin{eqnarray}
   h_c = \frac{D^2}{K-(X-D-1) \left\langle\sum_{\nu=1}^K 1/\lambda_\nu\right\rangle}.
   \label{hc2}
\end{eqnarray}
It is interesting that the correction in the denominator changes sign when $X=D+1$.

In the general case a closed form approximation can be obtained for small $K$. The limit spectral density of the nonzero eigenvalues of Wishart matrices is given by the Marcenko-Pastur law \cite{Marcenko1967}
\begin{eqnarray}
   \rho(\lambda) &=& \frac{1}{2\pi \min(D,X) }
               \sqrt{\frac{(\lambda-\lambda_{\min})
               \left(\lambda_{\max}-\lambda\right)}{\lambda^2}}\, ;
                    \nonumber\\
               \lambda_{\min} &=& \left(\sqrt{D}-\sqrt{X}\right)^2, \nonumber\\
               \lambda_{\max} &=& \left(\sqrt{D}+\sqrt{X}\right)^2,
\end{eqnarray}
which has a finite support $\lambda\in (\lambda_{\min},\lambda_{\max})$ and a square-root singularity as $\lambda\to \lambda_{\max}$. If $D,X$ is large and $K<<\min(D,X)$, the $K$ largest eigenvalues are all close to $\lambda_{\max}$. Thus in this limit
\begin{equation}
    \left\langle\sum_{\nu=1}^K \frac{1}{\lambda_\nu} \right\rangle \approx \frac{K}{\lambda_{\max}},
\end{equation}
which using Eq.\ (\ref{hc2}) in the limit $K<<D,X\to\infty$ gives
Eq.\ (\ref{hc2text}) in the text.

%\bibliography{fathbib}% Produces the bibliography via BibTeX.

\end{document}